\documentclass[twocolumn, times, tighten,twocolappendix]{aastex631}

\usepackage{graphicx,multirow,hyperref,url,color,xspace}
\usepackage{amsmath,amssymb}

\begin{document}

\title{Improved model of X-ray emission from hot accretion flows}
\shorttitle{X-rays from hot accretion flows}

\author{Andrzej Nied{\'z}wiecki}
\affiliation{Faculty of Physics and Applied Informatics, {\L}{\'o}d{\'z} University, Pomorska 149/153, 90-236 {\L}{\'o}d{\'z}, Poland; \href{mailto:andrzej.m.niedzwiecki@gmail.com}{andrzej.m.niedzwiecki@gmail.com}}
\author{Micha{\l} Szanecki}
\affiliation{Faculty of Physics and Applied Informatics, {\L}{\'o}d{\'z} University, Pomorska 149/153, 90-236 {\L}{\'o}d{\'z}, Poland; \href{mailto:andrzej.m.niedzwiecki@gmail.com}{andrzej.m.niedzwiecki@gmail.com}}
\author{Andrzej A. Zdziarski}
\affiliation{Nicolaus Copernicus Astronomical Center, Polish Academy of Sciences, Bartycka 18, PL-00-716 Warszawa, Poland;}
\author{Fu-Guo Xie}
\affiliation{Key Laboratory for Research in Galaxies and Cosmology, Shanghai Astronomical Observatory, Chinese Academy of Sciences, 80 Nandan Road, Shanghai 200030, People's Republic of China;}

\shortauthors{Nied{\'z}wiecki et al.}

\begin{abstract}
We have developed an improved model of X-ray emission from optically thin, two-temperature accretion flows, \texttt{kerrflow}, using an exact Monte Carlo treatment of global Comptonization as well as with a fully general relativistic description of both the radiative and hydrodynamic processes. It also includes pion-decay electrons, whose synchrotron emission dominates the seed photons yield at $\dot M/\dot M_{\rm Edd} \ga 0.1$ in flows around supermassive black holes. We consider in detail the dependence of the model spectra on the black hole spin, the electron heating efficiency, the plasma magnetization and the accretion rate, and we discuss feasibility of constraining these parameters by analyzing X-ray spectra of nearby low-luminosity active galactic nuclei. We note some degeneracies which hinder precise estimations of these parameters when individual X-ray spectra are analyzed. These degeneracies are eliminated when several spectra from a given source are fitted jointly, which then allows us to reliably measure the model parameters. We find significant differences with previous spectral models of hot-flow emission, related with the computational methods for Comptonization.  Finally, we briefly consider and discuss the dependence on the viscosity parameter and on the outflow strength.
\end{abstract}

\section{Introduction}
\label{sec:intro}

Optically thin, hot accretion flows represent the most physically complete explanation of low-luminosity black hole systems \citep[e.g.][]{1995ApJ...452..710N,1997ApJ...489..865E,2014ARA&A..52..529Y}. In particular, it is widely accepted that nearby, low-luminosity active galactic nuclei (AGN) are powered by such flows and spectral models  of hot-flow emission have been applied to a number of these AGNs \citep[e.g.][]{1996ApJ...462..142L,1996MNRAS.283L.111R,2009ApJ...699..513L,2014MNRAS.438.2804N,2016MNRAS.463.2287X}. However, the applied models involve several approximations which significantly reduce their accuracy. These include (i) the use of a pseudo-Newtonian potential of \cite{1980A&A....88...23P}, which fails in the innermost region, where most of the gravitational energy is dissipated, (ii) local approximations of Comptonization, which appear to be particularly incorrect in optically thin flows, and (iii) the neglect of nonthermal particles. 

Studies of hot flows including nonthermal electrons are rare and have been  either based on one-zone approximations, rather than on hydrodynamical solutions \citep[e.g.][]{2009MNRAS.392..570M,2009ApJ...690L..97P,2009ApJ...698..293V,2011MNRAS.414.3330V}, or applied to observations in the radio band only \citep[e.g.][]{1998Natur.394..651M,2019MNRAS.490.4606B}. On the other hand, general-relativistic (GR) hot-flow models, corresponding to the Kerr metric, were studied in only a few works \citep[in particular][]{1996ApJ...471..762A,1998ApJ...504..419P}, focused on hydrodynamical rather than radiative properties. A more complete GR study was presented in \cite{2000ApJ...534..734M}, which, however, used an approximate description of the Comptonization process. As we find below, an accurate treatment of Comptonization turns out to be a more important correction than a proper GR description or inclusion of nonthermal electrons.

In our previous works \citep{2010MNRAS.403..170X,2012MNRAS.420.1195N,2014MNRAS.443.1733N,2015ApJ...799..217N} we developed the computational model improving the above weaknesses. Here we use it to set up a large grid of solutions, which then allows us to construct a model, \texttt{kerrflow}\footnote{the \texttt{xspec} implementation of our code can be downloaded at 
\url{wfis.uni.lodz.pl/kerrflow}}, for direct analysis of the observed X-ray spectra. The hot flow model has a large number of free parameters and it is not feasible to construct a precise model in the whole parameter space. We, therefore, use the fixed values of some parameters, but we check how this may affect applications of our model. In particular, we include only nonthermal electrons produced by charged-pions decay. They should always be present in the innermost part of the flow, where protons have energies above the threshold for pion production, and then models neglecting this effect are not self-consistent. As we find in \citet{2015ApJ...799..217N}, the inclusion of the pion-decay electrons is more important in flows surrounding supermassive black holes, where their synchrotron emission strongly magnifies the seed photon input and then strongly increases the Compton cooling rate. In flows around stellar-mass black holes, the thermal synchrotron emission is comparatively stronger (see Section \ref{sect:model}), furthermore, the pion-decay electrons emit the synchrotron radiation mostly in the hard X-ray range and its Comptonization is inefficient as a cooling process, making this effect relatively unimportant. Therefore, we focus here on applications to low-luminosity AGNs.

\section{Model}
\label{sect:model}

\begin{figure}
\centering
\includegraphics[width=8cm]{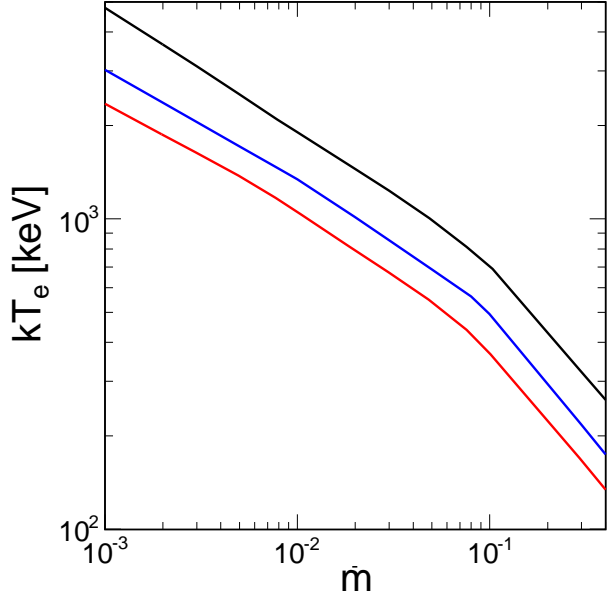}
\caption{Electron temperature in our solutions for $a=0.95$, $\delta=0.1$, $M = 8 \times 10^7 M_{\sun}$, and $\beta=1$ (red), 3 (blue) and 9 (black) from bottom to top. The figure shows $k T_{\rm e}$ at $r=3$, representing the region where the dominant contribution to the observed X-ray flux is produced.}
\label{fig:te}
\end{figure}

We briefly summarize here assumptions and computational methods of our model; detailed description has been presented in our previous works, referenced below. We consider a black hole, characterized by its mass, $M$, and angular momentum, $J$, surrounded by a geometrically thick accretion flow with an accretion rate, $\dot M$. We define the following dimensionless parameters: $r = R / R_{\rm g}$, $a = J / (c R_{\rm g} M)$, $\dot m = \dot M / \dot M_{\rm Edd}$, where $\dot M_{\rm Edd}= L_{\rm Edd}/c^2$, $R_{\rm g}=GM/c^2$ is the gravitational radius and $L_{\rm Edd} \equiv 4\pi GM m_{\rm p} c/\sigma_{\rm T}$ is the Eddington luminosity. We assume that  the density distribution is given by $\rho(R,z)=\rho(R,0) \exp(-z^2/2H^2)$, where $H$ is the scale height at $R$ and $z=R \cos \theta$. The ratio of the gas pressure (electron and ion) to the magnetic pressure is denoted by $\beta$; the fraction of the dissipated energy which heats directly electrons is denoted by $\delta$. We apply the usual assumption that the viscous stress is proportional to the total pressure, with the proportionality coefficient $\alpha$, and we assume $\alpha=0.3$; the dependence on $\alpha$ is discussed in Appendix \ref{sect:alpha}. We also assume that $\dot m$ does not depend on $r$; see Appendix \ref{sect:outfl} for the discussion of this assumption.

\begin{figure}
\centering
\includegraphics[width=8cm]{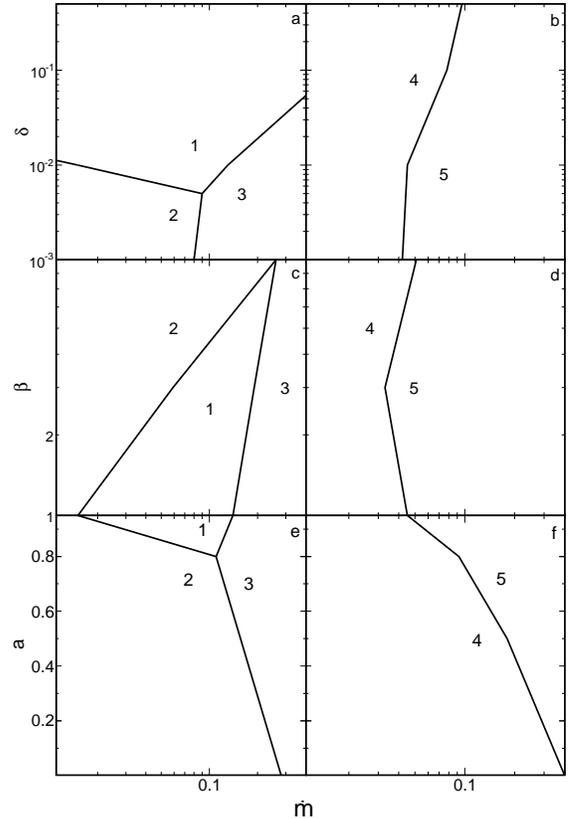}
\caption{Selected $\delta$-$\dot m$ (top), $\beta$-$\dot m$ (middle) and $a$-$\dot m$ (bottom) planes showing ranges of parameters, in which different physical processes dominate in our hot flow solutions for $M = 8 \times 10^7 M_{\sun}$. Panels in the left column show regions in which the total heating of electrons is dominated by (1) direct viscous heating, (2) compressive heating and (3) Coulomb transfer from protons. Panels in the right column show regions in which the cooling of electrons is dominated by thermal Comptonization of (4) thermal synchrotron photons and (5) nonthermal synchrotron photons produced by pion-decay electrons. (a,b) $\beta=1$ and $a=0.95$, (c,d) $\delta=0.01$ and $a=0.95$, (e,f) $\delta=0.01$ and $\beta=1$.}
\label{fig:process}
\end{figure}

\begin{figure*}
\includegraphics[width=17.4cm]{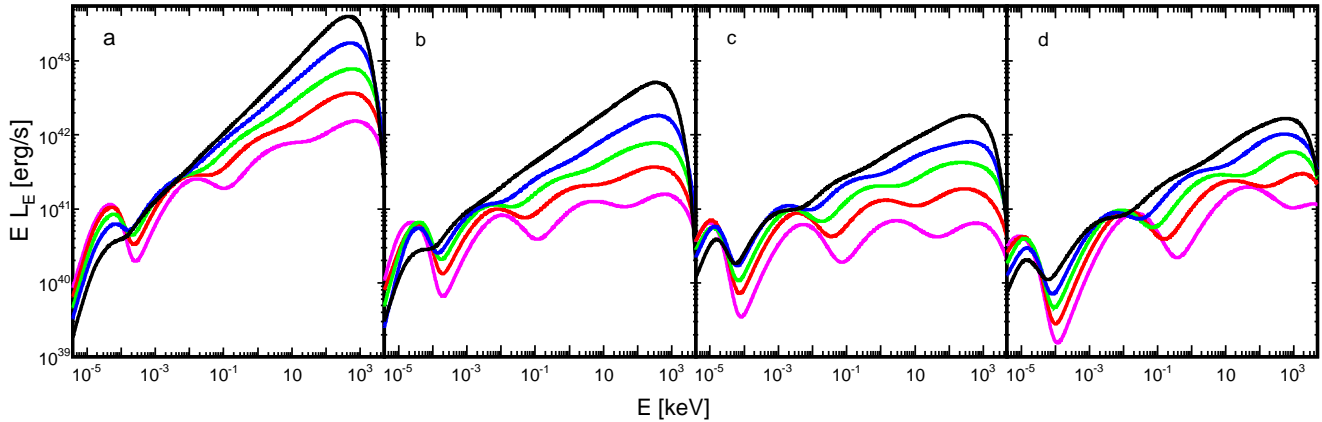}
\caption{Hot flow spectra produced by a hot flow for $M = 8 \times 10^7 M_{\sun}$ and (a) $\beta=1$, $\delta=0.5$, $a=0.95$, (b) $\beta=1$, $\delta=0.1$, $a=0.95$, (c) $\beta=1$, $\delta=0.1$, $a=0$, (d) $\beta=9$, $\delta=0.1$, $a=0.95$. In all panels $\dot m = 0.006$ (magenta), 0.0125 (red), 0.025 (green), 0.05 (blue), 0.1 (black) from bottom to top.}
\label{fig:spectra}
\end{figure*}

Our model includes a fully GR hydrodynamical description of the flow as well as an accurate treatment of the global Compton process taking into account all effects relevant for the Kerr metric. Our GR hydrodynamical model follows strictly \cite{2000ApJ...534..734M}.  Our modeling of Comptonization makes use of the GR Monte Carlo method of \cite{2005MNRAS.356..913N}, which follows \cite{1984AcA....34..141G} and extends it by including the special relativistic and gravitational effects affecting the photon transfer and energy gains in consecutive scatterings. We take into account only seed photons generated within the flow, neglecting any external irradiation. We consider the synchrotron and bremsstrahlung internal emission, however, the latter has a negligible effect for the range of parameters considered in our model. Furthermore, we take into account the presence of nonthermal electrons produced by the decay of charged pions, following \cite{2015ApJ...799..217N}. 

We strictly follow the procedure described in \cite{2012MNRAS.420.1195N}, extending it by the inclusion of nonthermal synchrotron emission. We find the self-consistent electron temperature distribution by iterating between the solutions of the electron energy equation  and the GR Monte Carlo Comptonization simulations until we find mutually consistent solutions. This procedure involves assumption that changes in $T_{\rm e}$ do not affect other parameters of the flow (density, proton temperature, scale height, velocity field), which limits the maximum luminosity of our solutions to $\sim 0.01 L_{\rm Edd}$.

Apart from the pion-decay $e^\pm$, we do not include additional nonthermal electrons, although these are likely present in hot flows \citep[e.g.][]{2016ApJ...822...88K,2019MNRAS.485..163K}. In some cases, such nonthermal electrons may strongly increase the synchrotron emissivity, e.g.\ by a factor of $\sim 10^5$ for a 1 per cent content of nonthermal electrons in a hybrid plasma with $kT_{\rm e} =50 $ keV \citep{2001MNRAS.325..963W}. We do not find this effect to be so extreme in our calculations. The thermal synchrotron emissivity is extremely sensitive to the electron temperature, $\propto T_{\rm e}^7$ \citep{1997ApJ...477..585M}. In our solutions, we typically find high electron temperatures at low $\dot m$, see Figure \ref{fig:te}, at which the thermal synchrotron emission would be outweighed by the nonthermal synchrotron emission only at a rather large energy content, $\ga 1$ per cent, of the nonthermal electrons. The electron temperature decreases with increasing $\dot m$, but this is also accompanied by (and partially due to) the increasing contribution of the synchrotron emission of pion-decay electrons, which again would be outweighed only for a rather large energy content, $\ga 1$ per cent, of the directly-accelerated, cooled electrons.

The normalization of the observed spectrum, $N$, is determined by $M$ and the distance, $d$. Regarding the former, all the relevant physical processes scale linearly with $M$, except for the thermal synchrotron emission whose luminosity is $\propto M^{1/2}$ \citep[cf.][]{1997ApJ...477..585M}. Furthermore, the frequency of the thermal synchrotron peak slightly increases with decreasing $M$. As a results, at low $\dot m$ -- when individual scattering bumps are visible in the spectrum -- the spectral shape is slightly dependent on $M$. We checked that differences in $M$ by a factor of several can be fully compensated by a change of $N$, $\propto M$, and a rather minor change of $\beta$ (e.g.\ by $\la 25$ per cent for the difference of $M$ by a factor of 2). However, differences in $M$ by orders of magnitude lead to significant spectral changes (due to the nonlinear scaling of the thermal synchrotron with $M$), in particular, for stellar-mass black holes the electron temperatures are significantly lower and the X-ray spectra significantly softer than for supermassive black holes \citep[see e.g.\ figure 2 and 3 in][]{2014MNRAS.443.1733N}.

We considered $M = 4  \times 10^7 M_{\sun}$ and $M = 8 \times 10^7 M_{\sun}$, and for each $M$ we found 312 solutions allowing us to construct the table model in \texttt{xspec} \citep{1996ASPC..101...17A}, in which we interpolate between various parameters to compute spectra for $10^{-3} \le \dot m \le 0.4$, $10^{-3} \le \delta \le 0.5$, $1 \le \beta \le 9$ and $0 \le a \le 0.95$. Our set of solutions also allows us to investigate the parameter space, in particular, to point out the range of parameters where the relevant physical processes dominate the total heating and cooling of the flow, as illustrated for selected ranges of parameters in Figure \ref{fig:process}. Note, in particular, that the decrease of $\beta$ strongly increases the cooling rate \citep[due to the increase of magnetic field combined with the decrease of scale height, cf.][]{2014MNRAS.443.1733N}, which reduces the electron temperature, see Figure \ref{fig:te}, and then strongly reduces the compressive heating, see Figure \ref{fig:process}(c). Note also that the relative importance of the synchrotron emission of pion-decay electrons decreases with decreasing $\dot m$, because the pion production rate depends on $\dot m^2$, and then thermal synchrotron dominates the seed photons input at low $\dot m$, see Figure \ref{fig:process}(b,d,f). The pion production and its effect on the cooling rate depends also significantly on $a$ \citep[cf.][]{2015ApJ...799..217N}, see Figure \ref{fig:process}(f).

\begin{figure*}
\includegraphics[height=7.5cm]{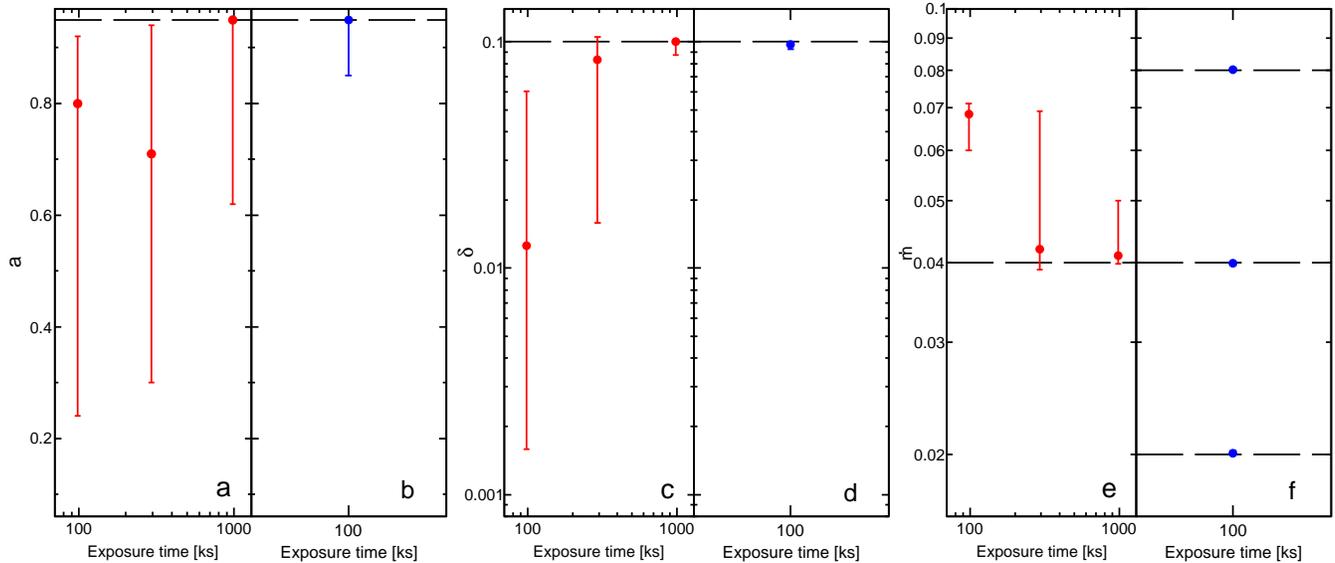}
\caption{Best-fit parameters inferred from the spectral analysis of synthetic  spectra for simultaneous {\it NICER} and {\it NuSTAR} observations of an AGN with $M = 8 \times 10^7 M_{\sun}$ at $d = 22$ Mpc. In all panels the dashed horizontal lines show the actual input parameters which were used to simulate the spectra. The red points in panels (a), (c) and (e) show parameters found by fitting individual spectra with different exposure times (100 ks, 300 ks and 1 Ms). The blue points in panels (b), (d) and (f) show parameters found by joint fitting of three spectra simulated for different values of $\dot m$ (given by the dashed lines in f), each for an exposure of 100 ks.}
\label{fig:fake}
\end{figure*}

\section{X-ray spectra}
\label{sec:result}

Figure \ref{fig:spectra} shows example spectra produced by hot flows. We see that the spectra exhibit a significant dependence on $\delta$, $a$ as well as $\beta$. We then check if these spectral differences allow us to estimate the model parameters by spectral modeling of the X-ray data from nearby AGNs.  We considered the case with $M = 8 \times 10^7 M_{\sun}$ and $d=22$ Mpc, which gives the X-ray photon flux ($\sim 10^{-11}$ erg s$^{-1}$ cm$^{-2}$ in the energy range 2--10 keV) representative for some low-luminosity AGNs. We simulated synthetic spectra for simultaneous observations by {\it NICER} and {\it NuSTAR}, using the {\tt fakeit} command in \texttt{xspec} and adding Poisson noise to mimic statistical fluctuations expected in real observations. We then jointly analyzed spectra generated in the 0.2--12 keV range for the XTI detector onboard {\it NICER} and in the 3--85 keV range for the FPMA and FPMB detectors onboard {\it NuSTAR}, which choice is suitable for our analysis as these detectors together provide good quality data in the photon energy range spanning over two decades of energy. We considered exposure times between 100 ks and 1 Ms. We have also considered datasets complemented by the {\it INTEGRAL}/ISGRI observations in the 20--250 keV range with the same exposure times, however, the results were similar to those for {\it NICER} and {\it NuSTAR} without {\it INTEGRAL}. While the exposure time of 1 Ms is not realistic, it allows us to  consider higher quality data which may be available with future instruments, in particular with {\it HEX-P}, with its planned improvement in sensitivity by over an order of magnitude over {\it NuSTAR} \citep{2019BAAS...51g.166M}.

We tested a number of such synthetic spectra and representative examples of fitting results are shown in Figure \ref{fig:fake}. In all cases the quality of the fits is good, with the reduced $\chi^2 \simeq 1$ for every best-fit. We found that fitting an individual spectrum does not allow us to reliably measure the model parameters, except for the (currently unrealistic) cases of exposure approaching 1 Ms, for which the fitted parameters are relatively close to their correct values, although still with some discrepancies. For shorter exposure times, the best-fitted values deviate significantly from the input values and their uncertainties are very large. This behavior is due to Poisson fluctuations (which are significant above 10 keV in spectra with the shorter exposure times) combined with degeneracies between model parameters. In particular, we note a degeneracy between $N$ and other parameters. Then, we found that these degeneracies are substantially reduced when several spectra from a given source are fitted jointly. We assumed that the source exhibits variability driven by the change of $\dot m$. In such a case, measurements of several spectra corresponding to different flux levels, with a realistic exposure of 100 ks, allow us to reliably estimate the values of $\dot m$, $\delta$ and $a$, see panels (b,d,f) in Figure \ref{fig:fake}, as well as the fitted $\beta = 3.00^{+0.03}_{-0.05}$ matches the input $\beta=3$, and $N=1.02^{+0.02}_{-0.03}$ (i.e.\ we recovered the actual normalization corresponding to the assumed $M$ and $d$).

The electron temperature predicted by the hot flow model is high and sensitive on the model parameters, see Figure \ref{fig:te}. The high-energy cut-offs of the corresponding Comptonization spectra occur at photon energies of hundreds of keV to several MeV. While testing this model prediction with present X/$\gamma$-ray facilities is not possible, it may be within the reach of future soft-MeV missions like {\it COSI}.

\begin{figure}
\includegraphics[width=8.4cm]{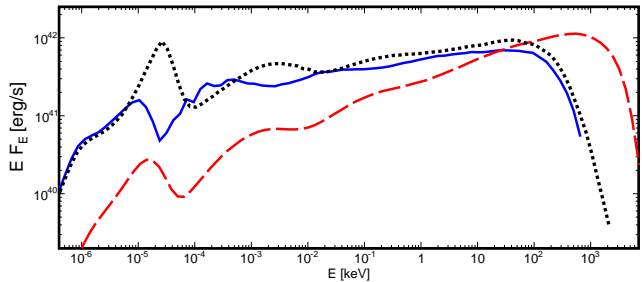}
\caption{Comparison of the \texttt{kerrflow} (dashed red), \texttt{riaf-sed} (solid blue) and \texttt{PNhot} (dotted black) spectra for $M = 8 \times 10^7 M_{\sun}$, $\beta=3$, $\delta=0.1$, $\alpha=0.3$, neglecting wind (see Appendix \ref{sect:outfl}). In \texttt{kerrflow}, $a=0$ and $\dot m = 0.1$.  In \texttt{riaf-sed} and \texttt{PNhot}, $\dot m = 0.01$ (accounting for the difference in definitions of $\dot m$ between these two models and  \texttt{kerrflow}).}
\label{fig:riafsed}
\end{figure}

Finally, we checked how strongly our model deviates from the non-GR models with local approximation for Comptonization, which have been applied to some low-luminosity systems.  We compared the \texttt{kerrflow} spectra with spectra computed using the \texttt{riaf-sed} model\footnote{publicly available at \url{https://github.com/rsnemmen/riaf-sed}} of \cite{2005ApJ...620..905Y,2007ApJ...659..541Y,2014MNRAS.438.2804N} and the similar model of \cite{2016MNRAS.463.2287X}, referred to here as \texttt{PNhot}. These two models adopt the pseudo-Newtonian gravitational potential \citep{1980A&A....88...23P} to mimic the effective potential of a Schwarzschild black hole, therefore, for the comparison we set $a=0$ in \texttt{kerrflow}. We find that for the same sets of parameters\footnote{we note that both \texttt{riaf-sed} and \cite{2016MNRAS.463.2287X} use the definition of $\dot M_{\rm Edd}$ including the 10\% radiative efficiency, so the factor of 10 should be applied for mapping the $\dot m$ parameter between these models and \texttt{kerrflow}}, the two non-GR models predict much softer spectra than \texttt{kerrflow}, see Figure \ref{fig:riafsed}. We suspect that this is due to computational problems in the treatment of Comptonization in \texttt{riaf-sed} and \texttt{PNhot}, as their spectra shown in Figure \ref{fig:riafsed} are clearly incorrect for $kT_{\rm e} \simeq (700 - 1000)$ keV found with these models in the whole region of X-ray production (i.e. within $100 R_{\rm g}$). In particular, the energy of the high-energy  roll-off is much too low. This weakness is most likely related with the use of the approximate formulae for Compton scattering of \cite{1990MNRAS.245..453C}.

We note also that the observed synchrotron radiation in {\tt kerrflow} is about an order of magnitude weaker than in the non-GR models with local Comptonization, due to a combination of two effects. First, a large part of the synchrotron emission from $r < 10$ is collimated toward the black hole and captured, which effect is included only in {\tt kerrflow}. Second, transfer of seed photons from smaller $r$ increases the Compton cooling rate and then reduces the temperature at $r \sim 100$. This effect (again taken into account only in {\tt kerrflow})  significantly reduces the synchrotron emission from this part of the flow \citep[cf. fig.\ 3ad in ][]{2010MNRAS.403..170X} which, for the models in Figure \ref{fig:riafsed}, contributes mostly in the millimeter range.

The difference between the spectra of the two non-GR models is mostly due to different corrections on radial velocity adopted in the region where the pseudo-Newtonian potential gives the velocity exceeding the speed of light.

\section{Summary and discussion}

We have developed a new \texttt{xspec} model for X-ray emission from hot accretion flows, \texttt{kerrflow}, which can be applied to low-luminosity black-hole systems to estimate the accretion rate, electron heating efficiency, black hole spin and plasma magnetization. We found that these parameters can be reliably determined in nearby AGNs, but this requires either the data quality significantly surpassing this currently available, or measurements of several spectra representing some variability of the source. Concerning the former, the planned {\it HEX-P} mission would offer an excellent prospect for testing the model predictions and estimation of its parameters, especially if combined with {\it ATHENA} \citep{2013arXiv1306.2307N} observations in the soft X-ray range and/or {\it COSI} \citep{2019BAAS...51g..98T} observations in the soft $\gamma$-ray range.  

We have compared our model with models previously applied to  black-hole systems and we found large discrepancies related with an inaccurate treatment of Comptonization in the latter. The computation of Comptonization is a relatively difficult issue. In our model we apply the Monte Carlo method, which allows us to precisely compute the spectrum (see also \cite{2014MNRAS.443.1733N}  for its comparison with an accurate semi-analytic model \texttt{compps} of \citet{1996ApJ...470..249P}). However, very time-consuming computations are required in this method (especially in its GR implementation), which is its major deficiency. The accurate method for computing Comptonization turns out to be the major improvement of our model, outweighing the other improvements, in particular the proper GR description.

The fully GR hot-flow model has been previously applied only to systems accreting at $\dot m \la 10^{-3}$ \citep[e.g.][]{2009ApJ...699..513L}, i.e.\ below the range of $\dot m$ considered in our model. At these low $\dot m$, the thermal Comptonization ceases to be a dominant cooling process and then a proper consideration of the non-local nature of this process is not crucial for the self-consistent solution in the central region. Still, an accurate Comptonization method would be important for reliable estimation of the X-ray spectrum and its comparison with observations.

Due to a large computational cost, construction of a precise model in the full parameter space is not feasible. Then, we fixed some parameters, but we have checked how this affects our results. In particular, we assumed a fixed, large viscosity parameter, $\alpha=0.3$. If the real $\alpha$ is lower, then application of our model underestimates the electron heating parameter, but this reflects the actual decrease of the viscous dissipation rate. We also neglected wind (see Appendix \ref{sect:outfl}), as we found no evidence, either observational or theoretical, for its presence at the site of X-ray production. We found that if winds are present in this region, parameters obtained with our model still give a reasonable approximation of the real parameters representing this region. We emphasize that we regard this approach to be valid only for investigation of the high-energy emission. Emission at low frequencies, especially in the radio band, is produced at larger distances and a proper incorporation of the wind is crucial for its estimation.

We also neglected nonthermal electrons which may be produced by direct acceleration processes. This is possibly the major shortcoming of our model. We have checked, however, that these nonthermal electrons would be important only for a rather large energy content in the electron distribution, $\ga 1$ per cent, which probably exceeds the level expected at the very efficient thermalization of electrons acting in hot flows for the accretion rates considered here \citep{1997ApJ...490..605M,2009MNRAS.392..570M,2011MNRAS.414.3330V}. Observational studies of hybrid plasmas in black-hole systems indicate that the energy content of nonthermal electrons may slightly exceed 1 per cent \citep[e.g.][]{2021ApJ...914L...5Z}, but these are only available for luminous (more than $ 0.01 L_{\rm Edd}$) hard states of black-hole binaries, for which the nature of the X-ray source is relatively uncertain, furthermore, kinetic simulations indicate that the content of nonthermal electrons decreases with decreasing luminosity \citep[cf. figures 1 and 3 in][]{2009MNRAS.392..570M}.

Finally, we considered only the internal emission from the flow, neglecting additional components, in particular emission from the jet, whose contribution is expected theoretically \citep[e.g.][]{2016ApJ...819...95O} and which may be indeed needed to explain some X-ray observations \citep[e.g.][]{2014MNRAS.438.2804N,2017MNRAS.468..435V}. Such a jet component is usually phenomenologically added to the theoretical X-ray spectra \citep[e.g.][]{2005ApJ...620..905Y}. Such phenomenological components can be easily incorporated in our model, yet we  deem comparisons of a precise, physical model of internal flow emission to be crucial for estimation of the range of parameters in which such extensions of the model are indeed needed.

\section*{ACKNOWLEDGMENTS}

This research has been supported in part by the Polish National Science Centre grants 2015/18/A/ST9/00746, 2016/21/B/ST9/02388
and 2019/35/B/ST9/03944. A.A.Z.\ and A.N.\ are members of International Team at the International Space Science Institute (ISSI), Bern, Switzerland, and thank ISSI for the support during the meetings. FGX is supported in part by National SKA Program of China (No.\ 2020SKA0110102), the National Natural Science Foundation of China (NSFC No.\ 11873074), and the Youth Innovation Promotion Association of CAS (Y202064).

\appendix

\section{Wind}
\label{sect:outfl}

The Bernoulli parameter of  hot flows  is positive  \citep[e.g.][]{1994ApJ...428L..13N} indicating that the hot accretion may generate winds. Estimations of their magnitude, however, cannot be performed analytically and must rely on  numerical simulations. These indeed indicate strong mass loss in winds \citep[although it was likely strongly overestimated in early works, see][]{2012MNRAS.426.3241N} and following \cite{1999MNRAS.310.1002S} it has become customary to assume a local accretion rate decreasing inward as a power-law,
\begin{equation}
\dot m = \dot m_{\rm out} (r/r_{\rm out})^s,
\label{eq:wind}
\end{equation}
at all radii down to the black-hole horizon, typically with $s \sim 0.5$ . However, more recent simulations with sufficient radial resolution \citep[e.g.][]{2012MNRAS.426.3241N,2015ApJ...804..101Y,2020ApJ...891...63W}, commonly find only a weak outflow at $r \la 100$ and no wind at $r \la 10$. Observationally, there is  evidence of winds in systems with very low accretion rates at radial scales of $r \sim 10^4 - 10^5$ \citep{2013Sci...341..981W,2019ApJ...871..257P}, but not at lower radii. Given the lack of any indications, either observational or theoretical, of a significant mass loss within the central $\sim 100 R_{\rm g}$, where the X-rays are produced, we do not include it in our model. Here we study inaccuracies which may potentially result from applications of the \texttt{kerrflow} model (without an outflow) to spectra produced at $s > 0$.

We considered two values of $s=0.2$ and $s=0.4$, and we simulated spectra for two values of the accretion rate for each $s$. Specifically, for both $s$ we used values of $\dot m_{\rm out}$ and $r_{\rm out}$, for which equation (\ref{eq:wind}) gives $\dot m_{10}  = 0.02$ or 0.03, where $\dot m_{10}$ is the local accretion rate at $r=10$ (which is the radial distance where most of the observed X-rays are produced at these values of $s$). We assumed that each spectrum was obtained simultaneously by {\it NICER} and {\it NuSTAR} with the same settings as in Section \ref{sec:result}. Then, we fitted jointly the two spectra linking the normalization parameter and allowing $\dot m$ to vary. The remaining parameters ($\beta_0=1$, $\delta_0=0.5$, $a_0=0.95$; subscript '0' denotes the actual value of the parameter used in the simulation) were not varied for the simulated spectra and they were linked in the fitting procedure. Except for the $\delta$ parameter, which was fitted at $\delta \simeq 0.3$ at $s=0.2$ and $\delta \simeq 0.4$ at $s=0.4$, we have recovered parameters close to their actual input values. The fitted accretion rate is $\dot m \simeq \dot m_{10}$, so using the $s=0$ model we properly recovered the accretion rate through the inner regions of the accretion flow. We emphasize, however, that this is true only if at least two spectra for a given $s$ are fitted jointly. Fitting these spectra separately gives some parameters differing from their input values by over an order of magnitude, especially for $s=0.4$, for which both simulated spectra where (individually) fitted with $\delta < 0.02$ (i.e.\ $< 0.04 \delta_0$) and with $\dot m \simeq 0.3$ (i.e.\ $\ga 10 \dot m_{10}$) and $N < 0.2$.    

\section{Viscosity parameter}
\label{sect:alpha}

Optically-thin accretion flows exist only below a critical accretion rate, $\dot M_{\rm crit} \sim \alpha^2 \dot M_{\rm Edd}$ \citep[e.g.][]{1996ApJ...462..136N,2012MNRAS.427.1580X}. If $\alpha$ is small, then optically-thin solutions are restricted to very low accretion rates and their luminosities are too low to explain the observed objects.  Therefore, applications of hot flows to real systems assume a large, fixed value of $\alpha$, typically $\alpha \ge 0.1$. We do the same in our \texttt{kerrflow} model, i.e.\ we use the fixed value of $\alpha=0.3$. Here we study related  inaccuracies, potentially resulting from applications of the \texttt{kerrflow} model (with $\alpha=0.3$) to spectra simulated for lower values of $\alpha$.

\begin{figure}
\centering
\includegraphics[width=8cm]{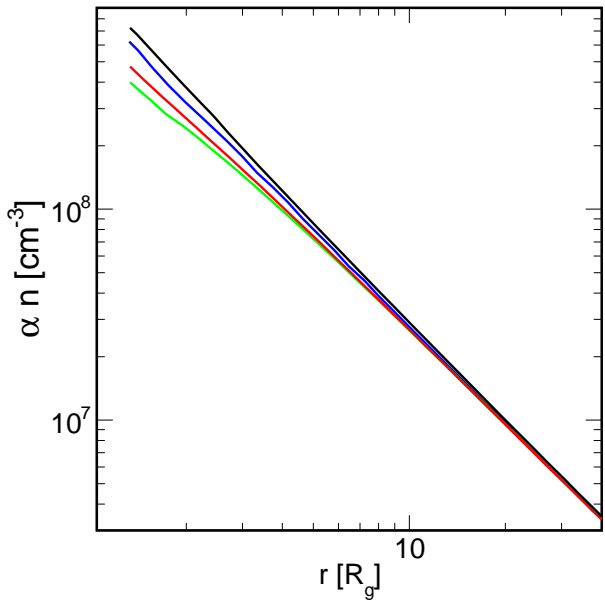}
\caption{Radial profiles of density for $\alpha=0.3$ (black), 0.2 (blue), 0.1 (red) and 0.07 (green) from top to bottom. The density is multiplied by the viscosity parameter to demonstrate departures from the self-similar scaling (i.e.\ $n \propto \alpha^{-1} r^{-1.5}$) at $r \la 10$ and their dependence on $\alpha$. $M = 8 \times 10^7 M_{\sun}$, $\dot m=0.01$, $a=0.95$, $\beta=1$, $\delta=0.1$.}
\label{fig:alpha}
\end{figure}

We considered two values of $\alpha=0.1$ and $\alpha=0.2$, and for each $\alpha$ we simulated spectra for three values of $\dot m_0 = 0.01$, 0.03 and 0.1 (in the following subscript '0' denotes the actual value of the parameter used in the simulation). We assumed that each spectrum was observed jointly by {\it NICER} and {\it NuSTAR} with the same settings as in Section \ref{sec:result}. Then, we fitted jointly the three spectra linking the normalization parameter and allowing $\dot m$ to vary. The remaining parameters ($\beta_0=1$, $\delta_0=0.1$, $a_0=0.95$) were not varied for the simulated spectra and they were linked in the fitting procedure. The major difference between the fitted and the actual values of the model parameters concerns $\delta$, which was fitted at $\simeq 0.6 \delta_0$ for $\alpha = 0.2$ and at $\simeq 0.2 \delta_0$ for $\alpha = 0.1$. The accretion rate parameter was fitted at $\simeq \dot m_0$ for $\alpha = 0.2$ and at $\simeq 2 \dot m_0$ for $\alpha = 0.1$, then, it does not follow $\dot m \propto 1/\alpha$, which we could expect guided by the self-similar scaling of density. The remaining parameters were fitted at their actual input values.

These fitting results properly represent the dependence of {\it global} hot-flow solutions on $\alpha$, inferred in a non-GR model by \cite{1997ApJ...476...49N} and confirmed with the GR description by \cite{1998ApJ...504..419P}. Firstly, the radial density profiles approach the self-similar form at $r \ga 10$, however, at $r < 10$ they exhibit a weaker dependence on $\alpha$ (see Figure \ref{fig:alpha}) as a results of the change of the pressure gradient in this region with changing $\alpha$ \citep[see figure 4 in][]{1997ApJ...476...49N}. This results in a weaker dependence of the $\dot m$ parameter on $\alpha$ than implied by the self-similar scaling. Secondly, the angular momentum profile flattens \citep[see figure 3 in][]{1997ApJ...476...49N} and hence the energy dissipation rate decreases with decreasing $\alpha$. Then, the fitted values of $\delta$, much lower than its real values in low-$\alpha$ models, reflect the actual reduction of the viscous heating rate. 
 
\bibliography{kerrflow}{}
\bibliographystyle{aasjournal}

\end{document}